# Extreme laser pulses for possible development of boron fusion power reactors for clean and lasting energy

H. Hora[*1], S. Eliezer[2], G. J. Kirchhoff[3], G.Korn[4], P. Lalousis[5], G. H. Miley[6] and S. Moustaizis[7]

[1]Department of Theoretical Physics, University of New South Wales, Sydney 2052, Australia
[2]SOREQ Research Centre, Yavne, Israel
[3]UJG Management GmbH, 85586 Poing, Germany
[4] ELI-Beam, Prague, Czech Republic.
[5]Institute of Electronic Structure and Lasers FORTH, Heraklion, Greece
[6]Dept. Nuclear, Plasma & Radiol. Engineering, University of Illinois, Urbana IL, USA
[7]Technical University Crete, Laboratory of Matter Structure and Laser Physics, Chania, Greece
*h.hora@unsw.edu.au

## ABSTRACT

Extreme laser pulses driving non-equilibrium processes in high density plasmas permit an increase of the fusion of hydrogen with the boron isotope 11 by nine orders of magnitude of the energy gains above the classical values. This is the result of initiating the reaction by non-thermal ultrahigh acceleration of plasma blocks by the nonlinear (ponderomotive) force of the laser field in addition to the avalanche reaction that has now been experimentally and theoretically manifested. The design of a very compact fusion power reactor is scheduled to produce then environmentally fully clean and inexhaustive generation of energy at profitably low costs. The reaction within a volume of cubic millimetres during a nanosecond can only be used for controlled power generation.

**Keywords:** boron laser fusion, ultrahigh acceleration, ultrahigh magnetic fields, avalanche HB11 reaction,

## 1. INTRODUCTION

The pollution of the atmosphere with carbon dioxide $CO_2$ was early recognized in a profoundly precise way by Al Gore [1] – when beginning his leadership in this field – with warning into what catastrophe the climate change will go by threatening the whole mankind [2]. The concentration of $CO_2$ in the atmosphere was for few thousand years on the level of 285 ppm (parts per million) keeping an ideal climate on earth by a green house effect but since about 1800, using fossil carbon for heating and for driving the steam engines and other power generators, it is now above 400 ppm. This substantial change can be seen from the thickness of tree rings showing the solar cycles and the assimilation of carbon in plants since few thousand years being constant, but in contrast now since 1960 steadily increasing to more than two times higher values (Fig. 5 of [3]). Melting the Greenland ice will increase the ocean level by 6 meters [3] and by more when adding the melting of parts of Antarctic ice.

The amount of using fossil carbon for the human civilization including the necessary use for the mobility in motor cars and other transport could well be acceptable at the level of about 1960 but the level has been increased by more than a factor 5 since, of which about 20% are the emission for the highly needed mobility by motor cars. These 20% could be saved if one would be able to completely stop the burning of tropical forests within a very short time for which the only attempt on a government level was tried in Australia [4].

Reducing the $CO_2$ emission into the atmosphere is progressing by using solar energy, wind craft or by about 10% from nuclear fission power reactors. The nuclear accidents at Three Mile Island, Chernobyl or Fukushima were exclusively caused through human mistakes caused by not obeying the well established rules of safety when intentionally switching off safety equipments or not building sufficiently high walls against tsunamis. The energy from the myriads of stars is mostly produced by nuclear fusion of hydrogen into helium and other light nuclei. To reproduce this for power generation on earth in a controlled way with power reactors is an important aim and enormous sums and most sophisticated exploration have been invested on earth but no energy producing fusion reactor has been achieved yet.

The key pioneer on plasma physics for this research, Edward Teller (see pages 2 to 4 of [5]), explained how difficult it is to solve this problem. These problems for the theory of high temperature plasma led Lord May of Oxford (R.M. May) to discover the fundamental theory of "complex systems" how to derive stabilizing properties [6] (see pages 2 to 3 of [5]). This also could be applied to other fields of science as animal populations or infectious diseases or even to the financial crisis [7].

A further basic aspect is the priciple of nonlinarity based on considertions of Richard Feynman (see Section 6.3 of [5]) initiated by research in laser interaction with plasmas [8]. This is related to well elaborated arguments by celebrities about the saturation or the end of physics while – in contrast – results of nonlinearity of laser-plasma interaction had just opened a door to a new dimension of physics.

This is just related to the discoveries of HB11 fusion for exploring a deviation from thermal equilibrium solutions and overcomming the limits of restrictions in linear physics [9] leading to the principle of nonlinearity [5]. This was crucial for opening a new option to overcome the climate problems by the following described discovery of a new kind of a power reactor with boron laser fusion.

This may be an example why a re-evaluation of the importance of nuclear energy may be recommended for solving the crucial problems of the dramatic climate change. It is positive that a Breakthrough Energy Coalition [10] is not only to support the usual solar energy or wind power generation but to provide a capacity for basically new alternatives or being aware that still undiscovered future innovations may change the whole scenario. One has not to miss the fact that a 100% carbon-free energy exists is in France or Sweden by using nuclear energy. This was formulated by the former Governor of South Australia, Admiral Kevin Scarce [11]. Just there it happened that a one-sided politologically oriented government based 40% of electricity generaton on expensive windcraft, until a rather normal storm had knocked down a number of these generators and the state had a black-out for several days. For the fission reactors, alternatives with small size module generators have been developed [12] and the most economic and least energy consuming isotope separation with lasers is availalbe on the market [13].

## 2. LOW DENSITIY VERSUS COMPACT BORON FUSION

The preferred reaction for fusion energy was from the beginning that of heavy with superheavy hydrogen, deuterium with tritium respectively, the DT reaction [14]

$$D + T = {}^4He + n + 17.4 \text{ MeV} \quad (1)$$

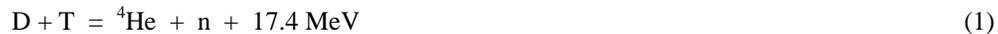

because it had the highest cross sections at the reaction energies below 100 keV compared with all other fusion reactions. The disadvantage of the reaction is the production of neutrons that are penetrating materials over large distances and when reacting then with stable, harmless nuclei changing these into dangerous radioactive waste [15]. About four times more neurons per gained energy are produced than from uranium and other fission reactions.

Neutron-free (aneutronic) fusion was discovered even before DT and measured [16] as reaction of the nuclei of the normal light hydrogen H, the protons p, with the boron isotope 11

$$H + {}^{11}B = 3\ {}^4He + 8.7 \text{ MeV} \quad (2)$$

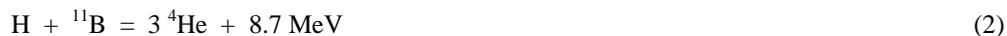

where the reaction energy was equally distributed as 2.9 MeV each to three harmless helium nuclei, called alpha particles, seen from cloud chamber pictures as Mercedes stars [17]. This HB11 reaction did not result in any neutrons primarily. Later studied small numbers of secondary reactions produces not more radioactivity per gained energy than burning coal due to its few parts of a millionth uranium in fossil carbon fuel. This emission is a completely ignorable amount of radioactivity.

The reaction cross sections of HB11 are so very much lower than for DT, that up to the knowledge of about 2009 the use for a fusion power reactor was many orders of magnitudes different. HB11 reactors were considered as absolute impossible. This was immediately evident since 1958 seen for reactors working with magnetic confinement under thermal equilibrium conditions as the stellarator or the tokamak. The generated cyclotron radiation emission was sufficiently low only for DT. All other fusion reactions had too high losses by radiation prohibiting BH11fusion at thermal equilibrium conditions due to the magnetic fields for confinement in tokamaks or stellarators. Working without magnetic fields by using lasers for compression, heating and thermal ignition of HB11, needs a compression of the fuel to more than 100,000times the solid state and an input laser pulse of more than 100 Megajoule to arrive at highest energy gains of less than 20 even by using best resonance conditions of the cross sections, alpha reheat and re-absorption of bremsstrahlung (Section 9.6 of [5]).

Any hope for a HB11 fusion [18][19] had to be based on thermal non-equilibrium, particle beam mechanisms, nonlinearities and further extreme conditions [9] for the scheme of Meglich [20] and others [21].

Only fusion reactions with thermal non-equilibrium or specially arranged ion beam conditions could be considered for a HB11 reactor. We shall differentiate between low density plasma reactions and those of compact designed with high reaction densities in the range of up to about the solid state density. Such a typical non-equilibrium plasma is known as Mather-Filipov plasma focus (see [22]) where a HB11 reactor has a chance to generate the necessary conditions within the category of low plasma density [23]. These comparably higher density plasma conditions than ITER etc. may finally lead to boron fusion but it should not be overlooked that the insider Robert Hirsch [24] underlined that "there is simply too much happening in these experiments to draw conclusions early on".

A further different scheme is studied by the Lockheed-Martin Cooperation (LM) known in order to produce non-thermal equilibrium HB11 plasma conditions for ion beam interactions. Not more than a patent application [25] has been disclosed for describing the proposed reactor and for understanding the properties. The plasma density is on the lower side as for other usual magnetic confinement.

Another non-equilibrium plasma scheme is the Hirsch-Miley [27] Inertial Electrostatic Confinement IEC experiment following a proposal by Farnsworth and envisaged by Peter Kapitza. A sphere S containing a low pressure gas with particles of nuclei for fusion reactions is separated from the inner wall by an isolated spherical grid G of more than -50 keV whose inner volume is of equal electric potential. The field between S and G results in a dark discharge in the low pressure gas. The ions of charge times 50keV energy passing the grid S and are then ballistic crossing the center of S colliding with other ions to produce fusion reactions by a field-free bouncing. Using DT gas, a dc emission can produce more than $10^{10}$ neurons per second for adding up the neutrons for criticality of fission reactors [27]. In case of an emergency, the neutron flux can be stopped instantly to shut down the reactor. The usually used neutron source from Californium-252 needs longer time to be removed and its frequent renewal is very expensive. The fusion reaction gain at the IEC can be increased by a factor between 10 and 100 [27] if the potential of S is not from a wired grid, but if there is a lamellar guiding by electrodes for better directing the ions towards the center. The final fusion gain is many orders of magnitude below break-even such that a power reactor could not be considered.

Instead of the spherical IEC device, a cylindrical section could be cut out (C-IEC), Fig. 1 [28]. The reacting ions are bouncing for and back within an electrically field free cylinder and are guided only by an external magnetic field. DT reactions have been measured similar to the spherical IEC indeed far below break-even for a fusion reactor. No HB11 reactions were measured yet.

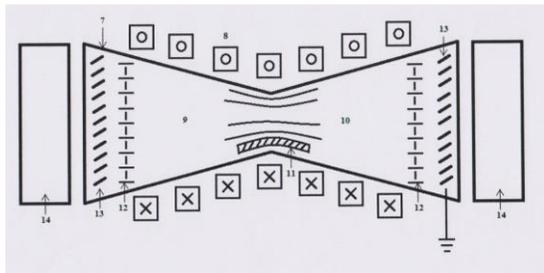

Fig. 1. Cylindrical modification of the Inertial Electrostatic Confinement (C-IEC) with explaining numbers see Patent Discolsure [28]

The scheme of a linear cylindrical ion beam bouncing reactions fulfils the beam reaction conditions known as Rostoker-Binderbauer (RB) device [29] based on RMF conditions and turbulence studies [30]. There is a similarity to the just described C-IEC with the cylindrical bouncing of the ion beams however where instead of the magnetic field coils in Fig. 1, a magnetic field as an "magnetic reversed field" configuration MRF is used. This cylindrical plasma has been studied on the basis of private investors to a very detailed developing level as Tri Alpha Energy TAE. Binderbauer realized – see the elaboration by Lev Grossman [32] – that the RB reactor within the work on fusion energy since the 1940's achieved progress "already almost done. It gets hyped to a level I think is very dangerous". The RB scheme was considered as an important step for going beyond DT in a qualitative hope towards HB11 fusion. To arrive at gains as explained by a high class analysis at an early stage [33] theoretically my reach a gain below break-even, but not more.

One step towards a more compact device was considered by extending the IEC device [18] with an improvement of the ion beam generation by extreme laser pulses [34]. Initially, the IEC following Hirsch used an injection of externally generated beams of ions with energies fitting good fusion reaction cross sections before working without extra ion injection and working only with the ions generated by dark ionization in the background gas due to the 50 Volts between the grid and the wall of the IEC. In order to provide higher reaction gains in the center of the IEC, the injection of comparably high energy ion beams are used by irradiating Petawatt-picosecond laser pulses generated from foils at the periphery of the IEC. Additional magnetic coils are to focus the ions into the center for higher density interaction. To the reported problems with overheating the guiding magnets a special geometry of the laser irradiated foils can be used for a ballistic focusing [35]. The plasma densities there are still of a lower level and not at the desired category of high plasma densities.

To summarize, any alternative HB11 fusion experiment (Tri Alpha, Plasma Focus, Inertial Confinement Fusion etc.) are of the category of lower densities at which research conditions for the experimental demonstration of fusion are aimed. But not any reaction has yet been measured. This is in contrast to high plasma density fusion by using laser irradiation, where the measurement of reactions did succeed. The historically very first thousand reactions were reported in 2005 [36], more than one million in 2013 [37] and one billion in 2014 [38]. This last experiment has been repeated with varying parameters and confirmed the extraordinary high HB11 fusion gains [39]. This is important for the following described compact fusion reactor.

**3. LASER DRIVEN PLASMA BLOCK IGNITION FOR BORON WITH MAGNEITC TRAPPING**

This paper distinguished boron fusion reaction of lower density plasmas from such of high densities for the design of the following boron laser fusion reactor. This is based on a sequence (about one shot per second) of reactions at solid sate fuel density in volumes of cubic millimeters of nanosecond duration with a spherical reactor size in the order of two or few meters more for delivering energy in the range of $100Million profit per year [40][41][42]**.

The essential difference between the laser interaction with a target between the use of laser pulses of nanosecond (ns) and that with picoseconds (ps) or shorter duration and very high power exceeding petawatt (PW) is essential for the new type of the HB11 reactor. The extensive laser-fusion studies during the last dozen years were mostly using ns pulses where the interaction processes were mostly dominated by thermal processes with thermalising the quiver energy of electrons in the laser field by classical (or quantum modified) collisions to heat ions delayed by the equilibration processes and that are possibly affected by instabilities and hydrodynamic motions to heat the ions whose pressure was then determining the subsequent ablation, compression and heating processes with nuclear ignition, re-absorption of bremsstrahlung, reheating by fusion reaction products including all the space and time variation of the optical constants. These difficulties were well described by Edward Teller (see [5]) also in view or the complex systems discovered by Lord May of Oxford (R. M. May [5]). As elaborated before, any consideration of HB11 fusion basically needs non-equilibrium plasma conditions [5][9]. This refers not only to plasmas and subsequent stabilization mechanisms but necessarily applies also to the principle of a non-thermal approach (Section 6.3 of [5]) in addition to the discovered complex systems.

The difference to the sub-ps interaction was revealed when attempting to explain the Linlor effect [43] why in addition to the thermally emitted ions of several eV energy from laser irradiated targets, ions with nearly thousand times higher energy were measured (see Fig. 2.5 of Ref. [5]). This goes back to the time where the dielectric plasma properties had to be clarified and their importance in the force densities given by Maxwell's stress tensor. At laser-plasma interaction, the force density **f** is not longer only determined by the gas dynamic pressure p but by the force $\mathbf{f}_{NL}$ due to electric **E** and magnetic **B** laser field of frequency ω [44],

$$\mathbf{f} = -\nabla p + \mathbf{f}_{NL} \qquad (3)$$

with the nonlinear force density due to the laser field

$$\mathbf{f}_{NL} = \nabla \bullet [\mathbf{EE} + \mathbf{HH} - 0.5(\mathbf{E}^2 + \mathbf{H}^2)\mathbf{1} + (1+(\partial/\partial t)/\omega)(\mathbf{n}^2-1)\mathbf{EE}]/(4\pi)$$
$$- (\partial/\partial t)\mathbf{E} \times \mathbf{H}/(4\pi c) \qquad (4)$$

using **1** as the unity tensor and **n** is the complex optical constant of the plasma given by the plasma frequency $\omega_p$. At plane laser wave interaction with a plane plasma front, the nonlinear force reduces to

$$\mathbf{f}_{NL} = -(\partial/\partial x)(\mathbf{E}^2+\mathbf{H}^2)/(8\pi) = -(\omega_p/\omega)^2(\partial/\partial x)(E_v^2/\mathbf{n})/(16\pi) \qquad (5)$$

showing how the force density is given by the negative gradient of the electromagnetic laser-field energy-density including the magnetic laser field from Maxwell's equations. $E_v$ is the amplitude of the electric laser field in vacuum after time averaging. The second expression in Eq. (5) is Kelvin's formulation of the ponderomotive force in electrostatics of 1845, indeed without magnetic fields, time dependence and the optical constant **n** of plasma. For general cases [5] all components of the stress tensor are needed otherwise the linear theory can arrive at completely wrong results as discussed with Feynman (see chapter 6.3 of [5]).

The basic and crucial difference between laser interaction by ns thermal interaction and that by ps non-thermal nonlinear force $\mathbf{f}_{NL}$ interaction is dominating in Eq. (3). For the ns interaction, the first term dominates at low laser intensities, while with ps, the second term dominates if the quiver energy of the electrons of the laser field is higher than their thermal energy of motion. A numerical example about nonlinear force acceleration of a slab of deuterium plasma irradiated by a neodymium glass laser pulse is shown in Fig 2. During the 1.5 ps, the plasma reached velocities above $10^9$ cm/s by the ultrahigh acceleration above $10^{20}$ cm/s$^2$. The generation of the plasma blocks, one moving against the laser light and the other into the higher density target, is the result of a non-thermal dielectric explosion and should not be misunderstood as radiation pressure absorption.

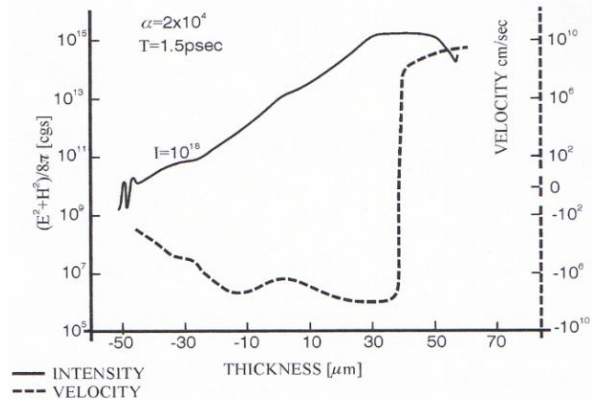

*Fig. 2. $10^{18}$ W/cm$^2$ neodymium glass laser incident from the right hand side on an initially 100 eV hot deuterium plasma slab whose initial density has a very low reflecting bi-Rayleigh profile, resulting in a laser energy density and a velocity distribution from plasma hydrodynamic computations at time t=1.5 ps of interaction. The driving nonlinear force is the negative of the energy density gradient of the laser field ($\mathbf{E}^2+\mathbf{H}^2$)/8π expressing the intensity. The dynamic development of temperature and density had accelerated the plasma block of about 20 vacuum wave length thickness of the dielectric enlarged skin layer moving against the laser (positive velocity) and another block into the plasma (negative velocity) showing ultrahigh >$10^{20}$cm/s$^2$ acceleration ([44] figures 10.17a&b).*

The experimental proof of the ultrahigh acceleration was possible [45] in full agreement with the theory, after laser pulses of higher than terawatt (TW) power and about ps duration were available only thanks to the Chirped Pulse Amplification CPA [46]. On top the pulses had to have an extreme contrast ratio to avoid relativistic self focussing [47] (see p. 180 of [5]). With these ps ultrahigh accelerations the plasma block ignition of solid density DT by the nonlinear force [48] was possible and updated using the one-fluid hydrodynamics by Chu [49] and Bobin [50] with many more details about the reaction over nanoseconds [51] and shock generation [52]. When the computations were performed with the binary fusion reaction cross sections of HB11 fuel instead of DT, a very surprising result was experienced. The ignition threshold for HB11 is at about the same value [53][54] as for DT at this non-thermal ps-laser block ignition **bridging the fife orders of differences** known from thermally dominated laser fusion processes [5][54] **. Further four orders of magnitudes** were achieved when instead of the pessimistic assumption of binary reaction the following described experimentally and theoretically proved avalanche reactions are used.

Up to this here considered point, all mentioned computations for the ps-laser pulse block ignition were for plane geometry laser pulses perpendicularly incident on a plane target. For a laser beam of circular cross section any lateral energy losses from the target have to be taken into account. For the aim of a complete trapping of the reaction cylinder below the circular interaction area in the irradiated fuel the use of a very high magnetic field parallel to the cylinder can be used for a complete isolation or trapping of the cylinder against energy losses. This is possible by using the now available ultrahigh magnetic fields of 4.5 kilotesla produced within the coils of the apparatus shown in Fig. 3 following the experiments by Fujioka et al. [55].

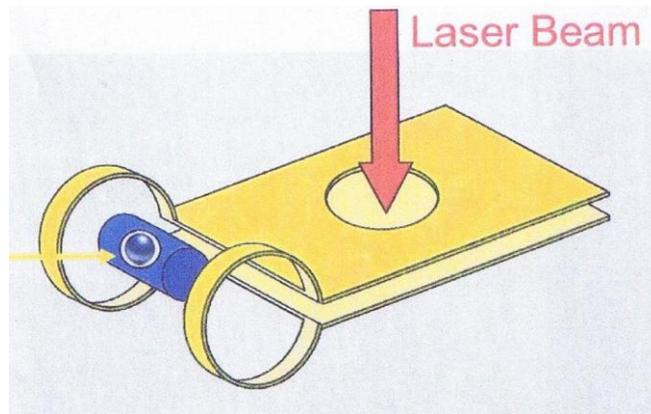

*Fig. 3. Using the coil for laser triggering of the pulsed current for generating of the magnetic field of 4.5 kiloesla generated in the loop up to values during more than a nanosecond (Fujioka et al. [55]. A cylindrical co-axial solid density HB11 fuel probe is located within the loop for end-on irradiation by a picosecond laser pulse to initiate the HB11 fusion reaction.*

Firing a laser beam of nanosecond duration and 1 kJ energy into the hole of the plates of Fig. 3 is triggering a current in the loops where a magnetic field of 4.5 kilotesla was measured [55]. Most of the energy of the laser pulse in Fig. 3 goes into charging the plates which energy – apart from some losses – is converted then into magnetic field energy in the coils during one to two nanoseconds. Placing into the coil a co-axial cylinder of solid density HB11 fuel of larger radius than the radius of a second picoseconds laser pulse results in a direct drive plasma block ignition of the fuel cylinder end-on. Hydrodynamic computations showed the plasma dynamics of the magnetically trapped or confining cylinder below the second laser pulse for the magnetic field of 4.5 and of 10 kilotesla at least for one nanosecond [40][56][57] (see also Figs. 10.13 to 10.23 of [5]).

## 4. BORON LASER FUSION REACTOR

Using the result to initiate a fusion reaction unit in the described way in the cylindrical fusion fuel of Fig. 3 by the picosecond laser pulse for a reaction within the 10 kilotesla magnetic field with trapping during more than one nanosecond duration, such a fusion reaction unit can be used in the center of a spherical reaction unit described in Fig. 4.

All these calculations are similar to the DT fusion using binary reactions without the secondary alpha avalanche reactions. The secondary reactions of the 2.9 MeV alphas when hitting a boron nucleus and transferring about 600 MeV energy by central collision are not included in the computations. The gyro radius of the alpha particles at 10 kilotesla magnetic fields is 42.5 μm and their mean free pass for collective stopping at solid state density is nearly independent on the electron temperature in the range of 60 μm at solid state density but is considerably larger in plasmas according to measurements at GSI Darmstadt [62][63] such that an avalanche multiplication is resulting in an exponential increase of the fusion gain until fuel depletion.

Estimations as for the cylindrical geometry [40] of the reaction unit of Fig. 3 show how a ps-30PW laser energy input into the block for the initiation of the flame of 30 kJ can produce alpha energy of >1 GJ. By this way, the requested fusion gain for DT of 10000 postulated by Nuckolls et al. for a power station [64] are then fulfilled. The aim to produce more than 100 MJ fusion energy per pulsed fusion shot was also underlined by Feder [65] mentioning Dawns Flicker [66]. Her understanding with respect to the costs for a fusion power station is evident. The scheme of Nuckolls et al [64] using relativistic electron beams for fast ignition arrives at comparable values for HB11. It is remarkable that the alpha-avalanche process is arriving at comparable values with clean HB11 fusion above those with DT.

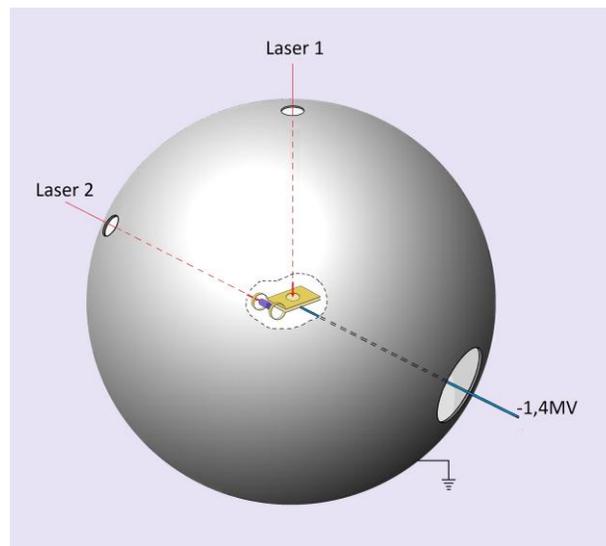

*Fig, 4. Scheme of an economic electric power reactor for production boron laser fusion, absolutely free from the problem of dangerous nuclear radiation [41][42][57] to [61]. Description of central reaction unit, see Fig. 3*

The energy of the alpha particles can be converted by more than 97% with a minimum of thermal energy losses when being slowed down by an electric field if the reaction unit is negatively charged at nearly 1.4 megavolts (MV) against the earthed reactor wall. The energy is given by the number of alpha particles times 2.8 MV which energy can be converted into three phase ac-electricity by using the HVDC (high voltage direct current) transmission technology used for electricity power transmission over 1000 km or much higher distance with minimum of losses [67][68]. If the reactor would work with one shot per second, the average dc current is 780 Amp at 1.4 MV.

The momentum of the alpha particle to the reactor wall by the alpha particles of GJ energy is reduced compared to the detonation of chemical explosives. This reduction is determined by the measures of the energetic particles. For chemical reactions the energy is up to the order of eV, while the energy of the particles of a nuclear reaction is up to the range of 10 MeV. The shock of the detonation is reduced by the square root of the ratio of the mentioned energies, i.e. in the range close to $3\times10^3$. The shock to the reactor wall is than comparable to an explosion in the range than 50g TNT, against which a >2 m diameter reactor wall of few cm thickness can sustain.

For an operation of up to one laser shot per second a feeding mechanism of the reaction units into the center of the reactor in vacuum at the voltage of -1.4 MeV is needed. The unit is destroyed at each shot. The hole in the reactor wall through which the unit is moved has to have a bent edge to reduce any vacuum discharge between the wall and the unit charged negative on1.4 MV together with the stick for guiding into the reactor center. The container of the feeding mechanism is on ground level like the reactor wall and has to have sufficient space for the moving stick, connected with the unit within vacuum and locks have to be provided for loading the units. This design is not trivial but well based on normal means. Lasers for the reactor for producing 30 kJ laser pulses of picosecond duration with a sequence of one shot per second are close to the present technology should be available within few years. At the moment [69] lasers with 10 kJ pulses of 0.17 ps duration and operation of one shot per minute are available or 1 kJ pulses of the same duration at 10 Hz emission.

## 5. AVALANCHE REACTION OF HB11

One point of importance for the working properties of the boron laser fusion reactor of Fig. 4 is the confirmation of the avalanche HB11 reaction in plasmas. The initial computations [40] used the long expected but not sufficiently confirmed amplification of the reaction by subsequent multiplication from the three generated alpha particles which themselves may generate subsequently a maximum of further three reactions etc. as avalanche. Simplified estimations [70] indicated this possibility more substantially than the earlier assumed qualitative expectations. The publication of the computations [40] was just at the same time as the generation of one billion HB11 reactions [39] by irradiating 500 Joule laser pulses of 200 ps duration on extremely boron enriched silicon targets used in semiconductor studies [71]. The comparison with fusion gains of DT [57][58] permitted [42] the proof of the avalanche mechanism at plane geometry [53][72] and was supported [73] by the evaluation of the elastic collisions and an subsequent reactions of HB11 within the background plasma of more than 10 eV temperature of solid state density [38][73][74][75].

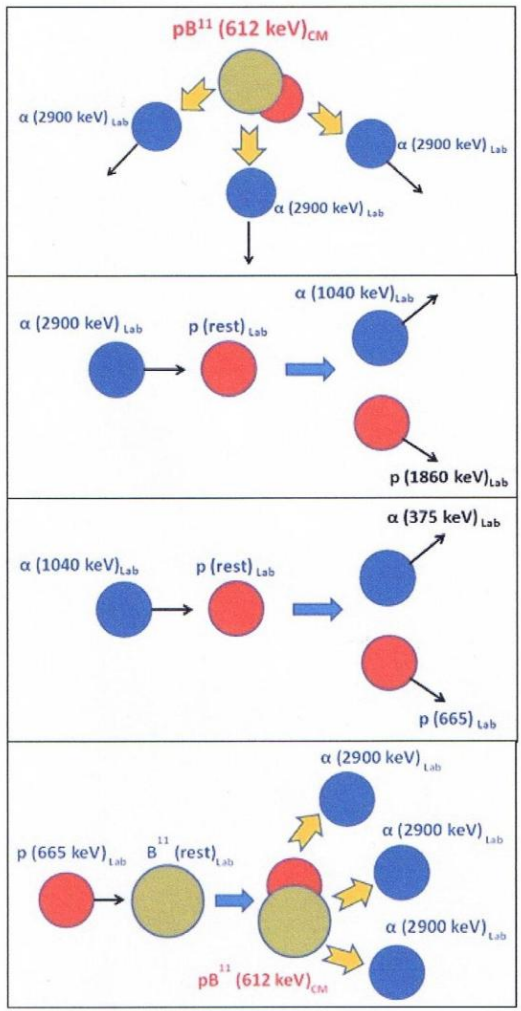

*Fig. 5: The avalanche scheme.*

A detailed explanation of the avalanche process is possible by the following evaluation of the elastic collisions [73] of the generated alpha particles. This is based on the force $\mathbf{f}_{NL}$ (Eq. 4) given by Maxwell's stress tensor as Lorentz and gauge invariant nonlinear force by quadratic terms of the force quantities $\mathbf{E}$ (electric field) and $\mathbf{H}$ (magnetic field) [48]. The non-linear force is dominating against the thermal forces [73], resulting in the acceleration of a plasma block causing non-thermal fusion ignition [38][42][60][61][74][76][77]. These alpha particles transfer energy in a broad energy range around 600 keV in the high density HB11 (noted as $pB^{11}$) plasma.

An initially resting $^{11}B$ or proton nucleus of mass $m_2$ gains energy from the energy $E_\alpha$ of an alpha particle of mass $m_1$. The maximum energy which can be acquired in the collision by a particle at rest is [75]

$$E_{2,final}(\text{initially at rest}) = \left[4m_1m_2/(m_1+m_2)^2\right]E_{1,initial}. \tag{6}$$

After a first collision of an alpha particle with a boron one gets:

$$E_{lab}(B^{11}) = \left(\frac{176}{225}\right)E_\alpha = 2,270[keV]; \quad E_{cm}(pB^{11}) = \left(\frac{1}{12}\right)\left(\frac{176}{225}\right)E_\alpha = 189[keV]$$

where $E_{lab}(B^{11})$ is the energy of the boron after the collision in the lab frame. While $E_{cm}(pB^{11})$, is the center-of-mass system energy of that boron and a proton at rest in the lab frame, and

$$E_{lab}(p) = \left(\frac{16}{25}\right)E_\alpha = 1860[keV]; \quad E_{cm}(pB^{11}) = \left(\frac{11}{12}\right)\left(\frac{16}{25}\right)E_\alpha = 1701[keV]$$

$E_{lab}(p)$ is the energy of the proton after the collision in the lab frame. $E_{cm}(pB^{11})$, is the center-of-mass system energy of that proton and a boron at rest in the lab frame.

After an alpha particle with an energy $E_\alpha = 2900$ keV has its second collision with a proton and this proton collides with a $^{11}B$ one gets in their center-of-mass system of reference an energy $E_{cm}(pB^{11})$

$$E_{cm}(pB^{11}) = \left(\frac{11}{12}\right)\left(\frac{16}{25}\right)\left(\frac{9}{25}\right)E_\alpha = 612.5[keV] \tag{7}$$

This energy is within the maximum cross section $\sigma_{max}$ of HB11 [78]. We get the energy for the HB11 maximum cross section from the alpha's collisions with protons that then collide with $B^{11}$ to get the fusion. We call this mechanism avalanche, because of the multiplication through generation of three secondary alpha particles by one primarily produced alpha particle. The avalanche scheme is described in Fig. 5. The alpha collisions with protons are more probable since the probability ratio is

$(n_p\sigma_{\alpha p}^{em}u_\alpha)/(n_B\sigma_{\alpha B}^{em}u_\alpha) \sim (n_p/n_B)[\,(1+m_\alpha/m_p)^2/(1+m_\alpha/m_B)^2][1/(Z_B^2)] \sim (1/2)\,(n_p/n_B)$

where the Rutherford cross section ($\sigma_{\alpha p}^{em}$, $\sigma_{\alpha B}^{em}$) has been used for the appropriate cross sections and $Z_B$ is the boron ionization degree. Since the hydrogen density is larger than the boron density by a factor of 10 we get the main chain of reactions as described in Fig. 5.

In this process we get 2 classes of proton densities, $n_{p1}$ that did not have any alpha collision and $n_{p2}$ that collided with alpha and got the right energy to have a p-$^{11}B$ collision at maximum nuclear cross section. It is conceivable to assume for this experiment [38] $n_p = n_{p1} + n_{p2}$ and $n_{p1} \gg n_{p2} = v_\alpha/3$ yielding the rate equation for the alpha particles

$$\frac{dn_\alpha}{dt} = 3n_{p1}n_B\langle\sigma v\rangle_T + 3n_{p1}n_B\langle\sigma v\rangle_{NT} + 3n_{p2}n_B\sigma_{max}u$$
$$\approx 3n_{p1}n_B\langle\sigma v\rangle_{NT} + n_\alpha n_B\sigma_{max}u. \tag{8}$$

The first term on the right hand side of equation (8) is given in reference [73] in order to explain the Prague experimental data [38]. However, this term is dependent on the ion temperature created in the laser plasma interaction.

$$<\sigma v>_T \left[\frac{cm^3}{s}\right] =$$
$$6.3820\times10^{-13}\zeta^{-5/6}\left(\frac{17.7080}{T^{1/3}}\right)^2 \exp\left(-\frac{53.1240}{T^{1/3}}\zeta^{1/3}\right) +$$
$$5.41\times10^{-15}T^{-3/2}\exp\left(-\frac{148}{T}\right)$$
$$\zeta = 1 + \frac{59.3570T - 1.0404T^2 + 9.1653\times10^{-3}T^3}{10^3 + 201.65T + 2.7621T^2 + 9.8305\times10^{-4}T^3}$$
ion temperature $T$ in keV.

In the Prague experiment, a total number of $N_{\alpha=}4\cdot10^8$ alpha particles per laser pulse were observed from a number of protons measured experimentally $N_H = n_p\Delta V = 10^{11}$ in a time $\Delta t = 10^{-9}$ s. The $N_\alpha$ observation of $N_{\alpha=}4\cdot10^8$ and calculation $N_\alpha = 3N_H n_B<\sigma v>_T\Delta t$, according to the first term of equation (8), limit the ion temperature to more than 100 keV. Therefore, according to the first term of equation (8) used in reference [74] to explain the data, the total expected fusion for an ion temperature of 100 keV is an order of magnitude less than the measured $N_\alpha$. Furthermore, we claim that for laser irradiance of $3\cdot10^{16}$ W/cm$^2$ reported in the Prague experiment an ion temperature of 100 keV is not conceivable.

The second term in equation (8) is a non-thermal equilibrium quantity that is related to the proton spectrum measured in reference [38]. The last term of equation (8) is caused by the protons that collided with the alphas and are returned back into the target by the inverted double layer (DL) simulations [76].

Taking the data from the Prague experiment, equation (8) can be solved numerically. In particular, the proton energy distribution as given in this experiment can be written as

$dN_p/dE = N_0$ [MeV$^{-1}$] for $0 < E < 1$ MeV and $dN_p/dE = 0$ for $E > 1$ MeV,

where $N_p$ is the proton volume integrated density number and $N_0 = 10^{11}$ is the total number of protons under consideration. This distribution implies

$$\frac{<\sigma v>_{NT}}{\sigma_{max}u} = \frac{\int_0^\infty f(E)\sigma(E)E^{1/2}dE / \int_0^\infty f(E)dE}{(1.2 barn)\sqrt{0.6 MeV}} \approx 0.40 \tag{9}$$
$$f(E) = \begin{cases} N_0 = 10^{11}[MeV^{-1}] \text{ for } 0<E<1MeV \\ 0 \text{ for } 1MeV<E \end{cases}$$

Therefore to a good approximation we get the following solution

$$N_\alpha = \frac{<\sigma v>_{NT}}{\sigma_{max}u}N_p\left(e^{\tau/\tau_A}-1\right) \approx 0.4N_0\left(\frac{\tau}{\tau_A}\right) \tag{10}$$
$$\tau_A \equiv \frac{1}{n_B\sigma_{max}u}$$

$N_0$ is of the order of few times $10^{11}$ and $N_\alpha$ of the order of $10^9$ are accordingly the volume integrated density numbers as given in the Prague experiment. $\tau_A$ is defined as the avalanche time and the interaction time $\tau$ to create alphas. In the Prague experiment $\tau_A$ is of the order of 100 ns ($n_B=10^{22}$ cm$^{-3}$, $\sigma_{max}=1.2$ barns and $u =10^9$ cm/s) which means that alphas are created during a couple of nanosecond.

The exponential term in our solution of equation (10) will be very large for $\tau \gg \tau_A$ in a pB11 fusion scheme as evaluated earlier in a summarizing way from the comparison with DT fusion in reference [42] by using a magnetic field to confine the laser produced plasma. In this case the avalanche process will dominate and therefore its application for a nuclear fusion reactor might be viable for the clean HB11 fusion. Computations [40][42][58,] done for cylindrical trapping with ultrahigh magnetic fields under the assumption of the avalanche, show that a 30 kJ laser pulse of ps duration could produce more than GJ energy in alpha particles. This laser energy is promising to achieve a HB11 fusion reactor. The measured strong elevation of the p-$^{11}$B fusion gain could only be explained as the result of the secondary p-$^{11}$B reactions caused by the avalanche process.

## 6. SUMMARY ABOUT THE BORON LASER FUSION REACTOR

Presently there are two leading developments where the clean HB11 fusion is favourable to the old and famous but not achievable yet in a reactor with DT fusion. These two new schemes are the "Tri Alpha" by Rostoker and Binderbauer [29][31] and our [42][73] "Avalanche" project [41]. The Tri alpha scheme is based mainly on the old magnetic confinement of classical reversed magnetic fields (RMF) with low density plasmas, while our high density plasma process is based on
(a) ultra-high plasma block acceleration by nonlinear forces at non-thermal direct laser-plasma target interaction as results of 1978 (see Fig. 2 with experimentally precise confirmation [45])
(b) in the presence of laser induced DC ultra-high magnetic fields [55], and
(c) by the discovery of the avalanche HB11 reaction [42][73] based on measurements [38][39][74].

Both developments are interesting but the question is what of them is more promising for the reactor concept. The "Avalanche" high plasma density project with combination of (a), (b) and (c) of these two models is the main issue in our route for clean nuclear fusion! Our Avalanche result has been **proven experimentally with an increase of nine orders of magnitude of alpha particle production** than expected from classical consideration. Our results can be verified with existing technologies and show that our clean fusion reactor can be achieved.

The result of the boron laser fusion reactor teaches how important it is to develop the theory of high temperature plasmas beyond the main stream of the thermal equilibrium state towards the plasmas with mixed states. These are the few hundreds keV non-thermalised energetic ions performing elastic collisions and fusion reactions in this broad energy range to happen within the plasma of about solid state densities and temperatures of 10 to 50 eV. Only this non-ideal plasma explains [73] the measured [37][38] exceptional high fusion reactions of HB11 with profound reconfirmation [39] as avalanche process [40]. The related studies [79] included the non-ideal plasmas for further interpretations.

The non-equilibrium and nonlinear states of plasmas were from the beginning crucial before 1987 [9] for HB11 fusion as a case related to conditions for Meglich's migma configuration [21] and related cases of the inertial electrostatic confinement [26]. The interrelation of problems of particle accelerators and plasma states is described as non-neutral plasmas, pioneered by Davidson [80]. The non-ideal plasmas for the avalanche processes [73] are extremely different from the initially studied cases of non-ideal plasmas [81] as given by dusty plasmas. After clarifying the wrong sign of the electric fields of the first considerations in view of the double layer state at plasma boundaries or surfaces, the dust plasmas were experimentally pioneered from the exhausts in the chemical plasmas in high power rockets by Fortov et al. where unique results were achieved in gravitation-free conditions in the space station [82]. These experiments were promoted by Fortov when he was Minister of Science of the Russian Federation and Deputy Minister President. In spite of this difference between the dust plasmas [81] and the avalanche HB11 fusion [73], the problems are related [79].

Finally it should be mentioned that the genuine hydrodynamic computation for HB11 laser fusion are going on to show their value [83] parallel to the wide spread PIC (particle in cell) computations where now the dielectric explosion (Fig. 2) is resulting [84][85][86] above the simplified radiation pressure push at low density plasmas, and supporting the result [86] how the dielectric effects due to inhomogeneous plasma densities increase the ion energies in nonlinear force driven plasma blocks at essentially higher densities [87]. Basic deviations from the three-alpha HB11 reaction could be excluded for very short-term interactions [88].

**Acknowledegement**. Great thanks are expressed to the numerous co-authors [41][42][51][52][54][58][60] [61][70][73][79][87] for their valuable contributions to the advancement towards the boron laser fusion reactor.

# REFERENCES

[1] Gore, Al, *Earth in the balance*. Plume Book, New York NY 1993
[2] Lovelock, James, *Gaia: A new look at the Life on Earth*, Oxford 2000
[3] Hora, Heinrich, *Climate Problems and Solutions* (in German) S. Roderer Publisher, Regensburg 2010
[4] Turnbull, Malcolm, Text of Law by Minister of Environment to the Australian Legislative, 20 March 2007
[5] Hora, Heinrich, *Laser Plasma Physics, SECOND EDITION* SPIE Press, Bellingham WA, 2016
[6] May, Robert M. Nature 238 (1972) 412
[7] Haldrane, A.G. and R.M. May, Nature, 469 (2011) 352
[8] Cicchitelli, L., H. Hora and R. Postle, Phys. Rev. A41 (1990) 3727
[9] Hora, H. Nuclear Instruments and Methods A, 271 (1988) 117
[10] Sweet, Cassandra, *Bill Gates' Breakthrough Energy Coalition*, The Wall Street Journal, Dec. 12, 2016.
[11] Scarce, Kevin, *A nuclear State in a Global Solution*, Four Societies Meeting, Sydney/AUS, 23 FEB 2017
[12] Paterson, Adrian, *Modular Nuclear Fission Rectors,* Four Societies Meeting, Sydney/AUS, 16 FEB 2015
[13] Goldsworthy, M.P., *The possibility for a commercial enrichment venture in Australia*, IAEA/INIS Reposi-tory No. 23042636; SILEX Technology Disclosure, 3 FEB 2017
[14] Oliphant, M.L.E., P. Harteck, and Lord Rutherford, Proc. R. Soc. London A 144 (1934) 692.
[15] Butler, D., 1996, Nature **380**, 655.
[16] Oliphant, M.L.E. and Lord Rutherford, Proc. R. Soc. London A 142 (1933) 259
[17] Oliphant, Sir Mark, showing cloud chamber pictures (when being Governor of South Australia 1974)
[18] Miley, G.H. *Fusion Energy Conversion* Hinsdale IL: Amer. Nucl. Soc. 1972
[19] Hora, H., *Laser Plasmas and Nuclear Energy* Plenum Press, New York 1975.
[20] Meglich, B., Nuclear Instruments and Methods A271 (1988) 13
[21] Maglich B and J. Morwood jr. Editorial: Aneutronic Power Princeton 1987. Nuclear Instruments and Methods A271 (1988) p. vii-viii
[22] Hora, H., R. Höpfl, and G.H. Miley. Plasma Physics and Controlled Fusion, 36, (1994) 1075
[23] Lerner, E. et al. Physics of Plasmas 19 (2012) 032704
[24] Anderson, M. IEEE Spectrum, 52, No 12, (2015) 9.
[25] McGuire, T. J, Lockheed-Martin US Patent Application 2014/0301517, Oct. 9, 204
[26] Hirsch, R. J. Appl. Phys. 38 (1972) 4522, see also: G.H. Miley & Krupakar Murali *Inertial confinement fusion (IEC)* Springer Heidelberg 2015
[27] Hora, H., M.A. Prelas and G.H. Miley. German Patent 4327920.1-34 granted 13 March 2003
[28] Hora, H. and G.H. Miley, German Patent Applic. 10118251.1 (2001)
[29] Rostoker N. and M. Binderbauer, US-Patent 6,664,740 B-2 (2002)
[30] Binderbauer, M.W. and N. Rostoker J. Plasma Physics 56 (1996) 451
[31] Rostoker, N., A. Qerushi and M. Binderbauer. Journal of Plasma Physics 22 (2003) 83
[32] Grossman Lev. Time Magazin Oct. 22, (2015)
[33] Wong, H.V. , B.N. Breizman, J.W. Van Dam. Tech. Rpt. Austin TX, US-DoD N00014-99-1-0888 (2001)
[34] Gruenwald, J., Plasma Physics and Controlled Fusion 59 (2017) 025011
[35] Moustaizis, S. (private commun.)
[36] Belyaev, V.S. et al. Phys. Rev. E 72 (2005) 026406
[37] Labaune, C., S. Deprierraux, S. Goyon, C. Loisel, G. Yahia & J. Rafelski. Communications 4, (2013) 2506
[38] Picciotto, A., D. Margarone, A. Velyhan, P. Bellini, J. Krasa, A. Szydlowski, G. Bertuccio, Y. Shi, A. Margarone, J. Prokupek, A. Malinowska, E. Krouski, J. Ullschmied, L. Laska, M. Kucharik, and G. Korn. Phys. Rev. X 4, (2014) 031030
[39] Giuffrida, L. This conference SPIE 10266-15
[40] Lalousis, P., H. Hora and S. Moustaizis, Laser and Particle Beams 32 (2014) 409.

[41] Hora H. and G.J. Kirchhoff, 2014 (23MAR) WIPO PCT Internat. Patent Disclosure (01 OCT 2015) WO 2015/144190 A1 PCT/EP2014/003281
[42] Hora, H., G. Korn, L. Giuffrida, D. Margarone, A. Picciotto, J.Krasa, K. Jungwirth, J. Ullschmied, P. Lalousis, S. Eliezer, G.H. Miley, S. Moustaizis and G. Mourou, Laser and Particle Beams. 33 (2015) 607
[43] Linlor, W.I., Appl. Phys. Lett. 3 (1963) 210.
[44] Hora, H. *Physics of Laser Driven Plasmas* (Wiley, New York 1981)
[45] Sauerbrey, R. 1996 Physics of Plasmas 3, (1996) 4712; I. Földes et al.Laser Physics 10, (2000) 264.
[46] Strickland, D. & G. Mourou, Opt. Commun. 56, (1985) 219.
[47] Zhang, M., J.T. He, D.B. Chen, Z.H. Li, Y. Zhang, Lang Wong., Z.H.B.F. Feng, D.F. Zhang, X.W. Tang & J. Zhang. Phys. Rev. E 57, (1998) 3745
[48] Hora, H. Phys. Fluids 12, (1969) 182
[49] Chu, M.S. Phys. Fluids 15, (1972) 412
[50] Bobin, J.L. Laser Interaction and Related Plasma Phenomena, H. Schwarz, H. Hora (Eds.), Vol. 3B, Plenum Press, New York, 1974, p. 465.
[51] Lalousis, P., H. Hora, S. Eliezer, J.-M. Martinez-Val, Moustaizis, G.H. Miley & G. Mourou. Phys. Letters A377, (2013) 885
[52] Hora, H., J. Badziak, M.N. Read, Y.-T. Li, T.-J. Liang, H. Liu, Z.-M. Sheng, J. Zhang, F. Osman, G.H. Miley, W. Zhang, X. He, H. Peng, F. Osman, S. Glowacz, S. Jablonski, S. Wolowski, Z. Skladanowski, K. Jungwirth, K. Rohlena & J. Ullschmied. Phys. Plasmas 14, (2007) 072701
[53] Hora, H. Laser & Part. Beams 27, (2009) 207
[54] Hora, H., G.H. Miley, M. Ghorannviss, H. Malekynia, N. Azizi & X-T. He. Energy and Environment Science 3, (2010) 479
[55] Fujioka S. et al. 2013 Scientific Reports 3, (2013) 1170
[56] Lalousis, P., S. Moustaizis, H. Hora and G.H. Miley, J. Fusion Energy 34 (2015) 62
[57] Hora, H. SPIE Newsroom 10.1117/2.121500.005965, see Google: "Boron Laser Fusion"
[58] Hora, Heinrich, Paraskevas Lalousis, Lorenzo Giuffrida, Daniele Margarone, Georg Korn, Shalom Eliezer, George H. Miley, Stavros Moustaizis and Gérard Mourou. SPIE Conf. Proc. 9515a, paper 9515-18 (2015)
[59] Moustaizis, S.D. et al. Proceed. SPIE 9515 (2015) 95151E
[60] Hora, H., P Lalousis, L Giuffrida, D Margarone, G Korn, S Eliezer, G H Miley, S Moustaizis, G Mourou and C P J Barty. Journal of Physics Conf. Proceed. 717 (2016) 012024
[61] G. H. Miley, H. Hora and G. Kirchhoff. Journal of Physics Conf. Proceed. 717 (2016) 012095
[62] Roth, M. et al., Abstract Fast Ignition Workshop, Oxford, September 15-18 (2014)
[63] Cayzac, W., V. Bagnoud, M.M. Basko, S.Bedacht, A. Balzevic, O. Deppert, A. Frank, D.D. Gericke, L. Hallo, A. Knetsch, D. Kraus, G.Maka, A. Ortner, K. Pepitone, G. Schaumann, T. Schlegel, D. Schumacher, A. Tauschwitz, J. Vorberger, F. Wagner, M. Roth, 2014 33rd ECLIM 2014 Conf. Paris, Abstracts p. 32.
[64] Nuckolls, J. L. and L. Wood *Future of Inertial Fusion Energy*, LLNL Preprint UCRL-JC-149860, Sept. 2002, (available to the public http://www.ntis.gov/); Future of Inertial Fusion Energy. *Proceedings International Conference on Nuclear Energy Systems ICNES Albuquerque, NM. 2002*, edited by T.A.Mehlhorn (Sandia National Labs., Albuquerque, NM) p.171-176.
[65] Feder, T. Physics Today 67, June 24 (2014)
[66] Flicker, D. Physics Today 67, June 25, see Feder (2014)
[67] Kanngiesser Karl-Werner., D. Hartmut Huang, Hans Peter Lips and Georg Wild. *HVDC Systems and their Planning Siemens EV HA 7*, Siemens Monographien, München (1994)
[68] Hammons, Thomas J., Victor F. Lesale, Karl Uecker, Markus Hausler, Dietmar Retzmann, Konstantin Staschus and Sebastian Lang, Proc. IEEE 100, No.2 (2012) 360 DOI: 10.1109/JPROC.2011.2152310
[69] Ditmire, T. This Conference, SPIE Proceedings 10241 (2017) 10241-25
[70] Hora, H., P. Lalousis, S. Moustaizis, I. Földes, G.H. Miley, X. Yang, X.T. He, S. Eliezer and J.-M. Martinez-Val. Shock Studies in Nonlinear Force Driven Laser Fusion with Ultrahigh Plasma Block Acceleration. *IAEA Proceedings Fusion Energy, San Diego Oct2012*. paper IFE/P6-03, 8 pages (IAEA Vienna 2013
[71] Hohenberger, M. et al. Physics of Plasmas 19, (2012) 056306
[72] Kline, J.L., S.A. Yi, A.N. Simakov, R.E. Olson, D.C. Wilson, G.A. Kyrala, T.S. Perry, S.H. Batha, E.L. Dewald, J.E. Ralph, D.J. Strozzi, A.G. MacPhee, D.A. Callahan, D. Hinkel, O.A. Hurricane, R.J. Leeper, A.B. Zylstra, R.R. Peterson, B.M. Haines, L. Yin, P.A. Bradley, R.C. Shah, T. Braun, J. Biener, B.J. Kozioziemski, J.D. Sater, M. M. Biener, A.V. Hamza, A. Nikroo, L.F. Berzak Hopkins, D. Ho, S. LePape, N.B. Meezan, D.S. Montgomery, W.S. Daughton, E.C. Merritt, T. Cardenas, E.S. Dodd. Developing one-dimensional implosions

for inertial confinement fusion science. High Power Laser Science and Engineering, 4 (2016) DOI: https://doi.org/10.1017/hpl.2016.43

[73] Eliezer, S., H. Hora, G. Korn, N. Nissim and J.M. Martinez-Val, Physics of Plasmas 23 (2016) 050704
[74] Margarone, D. et al. Plasma Phys. Contrl. Fusion 57 (2015) 014030
[75] Landau L.D., E.M. Lifshitz *Mechanics* 2$^{nd}$ Ed. Pergamon Press 1972, p. 17
[76] Hora, H., P. Lalousis, S. Eliezer. Physical Review Letters 54 (1984) 1650
[77] Mourou, G., C.P.J. Barty & M.D. Perry. Physics Today 51 No.1, (1998) 22
[78] Nevins, W.M. and P. Swain. Nucl. Fusion 40, (2000) 865
[79] Hoffmann, D.H.H., H. Hora, S. Eliezer, V.E. Fortov, N. Nissim, P. Lalousis, S. Moustaizis, J.-M.Martinez-Val. Conference Hirscheck/Germany, Jan. 2017
[80] Davidson, R.C. *Physics of Nonneutral Plasmas,* 2$^{nd}$ Edition, Imperial College Press London 2016
[81] Fortov, V.E. and T. Iakubov, *The Physics of Non-ideal Plasmas*, World Scientific Singapore 2002
[82] Fortov,V.E., *Extreme States of Matter*, Moscow 2009
[83] Moustaizis, S. et al. This conference Proc. SPIE 10241, paper 10241-55 (2017)
[84] Xu, Y., J. Wang, X. Qi, M. Li, Y. Xing and L.Yong, AIP Advances 6 (2016)105304
[85] Li, M., J.X. Wang, T. Yuan, Y.X. Xu, W.J. Zhu AIP Advances 7 (2017) 035007 (2017)
[86] Xu, Y., J. Wang, X. Qi, M. Li, Y. Xing, L. Long and W. Zhu, Physics of Plasmas 24, 033108 (2017)
[87] Yazdani, E., R. Sadighi-Bonabi, H. Afarideh, J.Yazdanpanah, H.Hora Laser & Part. Beams 32 (2014) 509
[88] Belyaev, V.S., V.P. Krainov et al. Laser Physics Letters 12 (2015) 096001